# Drug delivery with carbon nanotubes for in vivo cancer treatment

Zhuang Liu[1], Kai Chen[2], Corrine Davis[3], Sarah Sherlock[1], Qizhen Cao[2], Xiaoyuan Chen[2], Hongjie Dai[1*]

[1] Department of Chemistry, Stanford University, Stanford, CA 94305, USA

[2] The Molecular Imaging Program at Stanford (MIPS), Department of Radiology, Biophysics and Bio-X Program, Stanford University School of Medicine, Stanford, CA 94305, USA

[3] Department of Comparative Medicine, Stanford University School of Medicine, Stanford, CA 94305, USA

*Correspondence should be sent to: hdai@stanford.edu

**Chemically functionalized single-walled carbon nanotubes (SWNTs) have shown promise in tumor targeted accumulation in mice and exhibit biocompatibility, excretion and little toxicity. Here, we demonstrate in-vivo SWNT drug delivery for tumor suppression in mice. We conjugate paclitaxel (PTX), a widely used cancer chemotherapy drug to branched polyethylene-glycol (PEG) chains on SWNTs via a cleavable ester bond to obtain a water soluble SWNT-paclitaxel conjugate (SWNT-PTX). SWNT-PTX affords higher efficacy in suppressing tumor growth than clinical Taxol® in a murine 4T1 breast-cancer model, owing to prolonged blood circulation and 10-fold higher tumor PTX uptake by SWNT delivery likely through enhanced permeability and retention (EPR). Drug molecules carried into the reticuloendothelial system are released from SWNTs and excreted via biliary pathway without causing obvious toxic effects to normal organs. Thus, nanotube drug delivery is promising for high treatment efficacy and minimum side effects for future cancer therapy with low drug doses.**

# Introduction

A holy grail in cancer therapy is to deliver high doses of drug molecules to tumor sites for maximum treatment efficacy while minimizing side effects to normal organs.(1, 2) Through the enhanced permeability and retention (EPR) effect, nanostructured materials upon systemic injection can accumulate in tumor tissues by escaping through the abnormally leaky tumor blood vessels(3-6), making them useful for drug delivery applications. As a unique quasi one-dimensional (1D) material, SWNTs have been explored as novel drug delivery vehicles in vitro(7-9). SWNTs can effectively shuttle various bio-molecules into cells including drugs,(7-9) peptide,(10) proteins,(11) plasmid DNA(12) and small interfering RNA (siRNA)(13, 14) via endocytosis.(15) The intrinsic near infrared (NIR) light absorption property of carbon nanotubes has been used to destruct cancer cells in vitro(16) while their NIR photoluminescence property has been used for in vitro cell imaging and probing.(17) The ultra-high surface area of these 1D poly-aromatic macromolecules allows for efficient loading of chemotherapy drugs.(8) Various groups have investigated the in vivo behavior of carbon nanotubes in animals.(18-20) It is found that well PEGylated SWNTs intravenously injected into mice appear non-toxic over several months.(21) Nanotubes accumulated in the reticuloendothelial systems (RES) of mice are excreted gradually via the biliary pathway and end up in the feces.(22) Targeted tumor accumulation of SWNTs functionalized with targeting ligands RGD peptide or antibodies has shown high efficiency.(18, 20) These results set a foundation for further exploration of carbon nanotubes for therapeutic applications.

In the current work, we demonstrate SWNTs delivery of paclitaxel (PTX) into xenograft tumors in mice with higher tumor suppression efficacy than the clinical drug formulation Taxol®. The water insoluble PTX conjugated to PEGylated SWNTs exhibit high water solubility and maintain similar toxicity to cancer cells as Taxol® in vitro. SWNT-PTX affords much longer blood circulation time of PTX than that of Taxol® and PEGylated PTX, leading to high tumor uptake of the drug through EPR effect. The strong therapeutic efficacy of SWNT-PTX is shown by its ability to slow down tumor growth even at a low drug dose (5mg/kg of PTX). We observe higher tumor uptake of PTX and higher ratios of tumor to normal-organ PTX uptake for SWNT-PTX than Taxol® and PEGylated PTX, highly desired for higher treatment efficacy and lower side effect. PTX carried into RES organs by SWNT-PTX is released from the nanotube carriers likely via in vivo ester cleavage and are cleared out from the body via the biliary pathway. The non-cremophor composition in our SWNT-PTX, rapid clearance of drugs from RES organs, higher ratios of tumor to normal organ drug uptakes, and the fact that tumor suppression efficacy can be reached at low injected drug dose make carbon nanotube drug delivery a very promising nano-platform for future cancer therapeutics.

## Methods

**Functionalization of SWNTs with phospholipid – branched PEG**

One molar equivalent (eq.) DSPE-PEG5000-Amine (SUNBRIGHT® DSPE-050PA, NOF cooperation) was reacted with 5 eq. succinic anhydride in dichloromethylene ($CH_2Cl_2$, Aldrich) overnight at room temperature. After evaporating the solvent, the product was dissolved in water. The solution was dialyzed against water with a 3.5 kDa molecular weight cut off (MWCO) membrane for 2 days and then lyophilized into powder. The resulting DSPE-PEG5000-COOH was activated by 1.5 eq. dicyclohexylcarbodiimide (DCC, Aldrich) and 2 eq. hydroxybenzotriazole (HOBt, Aldrich) in $CH_2Cl_2$ at for 1 hour. 4 eq. 4-Arm-(PEG-Amine) (10kDa, P4AM-10, Sunbio) was added and the reaction solution was stirred for 2 days. After evaporating the solvent, water was added into the container and stirred for 1 hour. Solid precipitate (leftover DCC and HOBt) was removed by filtration via a 0.22 µm filter, yielding clear water solution of DSPE-PEG5000-4-Arm-(PEG-Amine) (see Fig.1a). The product was confirmed by MALDI (matrix-assisted laser desorption/ionization) mass spectrometry in Stanford PAN facility, showing no existence of starting DSPE-PEG5000 material. No further purification was performed since the excess hydrophilic 4-Arm-(PEG-Amine) molecules were confirmed to exhibit no binding affinity to nanotubes.

Raw Hipco SWNTs (0.2mg/mL) were sonicated in a 0.2mM solution of DSPE-PEG5000-

4-Arm-(PEG-Amine) for 30 min with a cup-horn sonicator followed by centrifugation at 24,000 g for 6 h, yielding a suspension of SWNTs with non-covalent phospholipid–branched PEG coating in the supernatant(14, 18, 23). Excess surfactant and un-reacted PEG molecules were removed by repeated filtration through a 100 kDa MWCO filter (Millipore) and extensive washing with water.

**Paclitaxel conjugation**

Paclitaxel (LC Laboratories) was modified by succinic anhydride (Aldrich) according to the literature, adding a carboxyl acid group on the molecule at the C'-2 OH position highlighted in Fig.1a.(24) 300nM of SWNTs (0.05mg/ml) with branched PEG-NH$_2$ functionalization was reacted with 0.3mM of the modified paclitaxel (dissolved in DMSO) in the presence of 5mM 1-ethyl-3(3-dimethylaminopropyl) carbodiimide hydrochloride (EDC, Aldrich) and 5mM N-hydroxysulfosuccinimide (Sulfo-NHS, Pierce). The solution was supplemented with 1x phosphate buffered saline (PBS) at pH 7.4. After 6 h reaction, the resulting SWNT-PTX was purified to remove un-conjugated PTX by filtration through 5 kDa MWCO filters and extensive washing.

UV-Vis-NIR absorbance spectra of the SWNT-PTX conjugates were measured by a Cary-6000i spectrophotometer. The concentration of SWNTs were determined by the absorbance at 808nm with a molar extinction co-efficient of $3.95 \times 10^6$ M·cm$^{-1}$ with an average tube length of ~100nm(16). Concentration of PTX loaded onto SWNTs was measured by the absorbance peak at 230 nm (characteristic of PTX, Fig.1a green curve, after subtracting the absorbance of SWNTs at that wavelength) with a molar extinction coefficient of $31.7 \times 10^5$ M·cm$^{-1}$. Note that thorough removal of free un-bound PTX was carried out by filtration prior to the measurement to accurately assess the amount of PTX loaded onto SWNTs. To confirm the PTX loading measured by UV-VIS, $^3$H-PTX (see the following paragraph) was conjugated to SWNTs. The PTX loading number on nanotubes measured by radioactivity was consistent to that measured by UV-VIS spectra, for same batches of samples.

PEGylated paclitaxel (PEG-PTX) was synthesized by reacting 1 eq. of 4-Arm-(PEG-Amine) (10 kDa) with 4 eq. succinic anhydride modified PTX in the presence of EDC/NHS at the same reaction condition as conjugation of SWNT-PTX. Excess unreacted PTX was removed by filtration via 5 kDa MWCO filters. The concentration of PEG-PTX was measured by its absorbance spectrum. In the case of radiolabeled $^3$H-PTX, 100 µCi (~5 µg) of $^3$H-paclitaxel (Moravek Biochemicals) was mixed with 10mg of regular non-radioactive paclitaxel and used for conjugation to obtain SWNT-PTX or PEG-PTX to impart radioactivity.

To make DSEP-PEG-PTX, as made DSPE-PEG5000-4-Arm-(PEG-Amine) (MW = 16 kDa) was purified by dialysis (membrane MWCO = 12-14 kDa) against water to remove excess 4-Arm-(PEG-Aimne) (MW = 10 kDa). Over 99% of unconjugated 4-Arm-(PEG-Aimne) was removed as confirmed by MALDI mass spectrum. The purified product was lyophilized (yield after dialysis ~ 50%) and stored at -20$^o$C. PTX conjugation was performed following the same procedure as described in the synthesis of PEG-PTX.

Taxol® was constituted following the clinical formulation. 6mg/ml of paclitaxel with or without addition of $^3$H-paclitaxel (50µCi/ml, ~2.5µg/ml) was dissolved in 1:1 (v/v) mixture of Cremophor EL (Aldrich) and anhydrous ethanol (Fisher) and stored at -20$^o$C.

**Cell toxicity assay**

4T1 murine breast cancer cell line (from American Type Culture Collection, ATCC) was cultured in the standard medium. Cells were plated in 96-wall plates and treated with different concentrations of SWNT-PTX, PEG-PTX or Taxol® for 3 days. Cell viability after various treatments was measured by the MTS assay with CellTiter96 kit (Promega).

**Animal model and treatment**

All animal experiments were performed under a protocol approved by Stanford's Administrative Panel on Laboratory Animal Care (APLAC). The 4T1 tumor models were generated by subcutaneous injection of $2 \times 10^6$ cells in 50 µl PBS into the right shoulder of female Balb/c mice. The mice were used for treatment when the tumor volume reached 50-100 mm$^3$ (~6 days after tumor inoculation). For the treatment, 150-200µl of different formulations of paclitaxel and SWNTs in saline was intravenously (IV) injected into mice via the tail vein every 6 days. The injected doses were normalized to be 5mg/kg of paclitaxel. The tumor sizes were measured by a caliper every the other day and calculated as the volume = (tumor length)$\times$ (tumor width)$^2$/2. Relative tumor volumes (Fig.2) were calculated as V/V$_0$ (V$_0$ was the tumor volume when the treatment was initiated).

**Tumor slice staining**

Tumors were placed into Optimal Cutting Temperature (OCT) medium immediately after being taken out from the mice, frozen by dry ice and stored at -80$^o$C. 5 µm thick tumor slices were cut by a Heidelburg microtome. The slices were stored at -80$^o$C until use.

**TUNEL staining.** Frozen tumor tissue slices from treated mice in storage were taken out from freezer and warmed for 20min at room temperature, then fluorescent TUNEL staining were conducted following manual instruction of In Situ Cell Death Detection kit (Roche, Indianapolis, IN ).

**Fluorescent staining of Ki67.** Frozen tissue slices from treated mice in storage were fixed with ice-cold acetone for 10 min and then dried for 30 min at room temperature, After 3x 5min rinse with PBS, slides were blocked with 10% goat serum in PBS for 15 min at room temperature. The slices were then incubated with rabbit anti-mouse Ki67 antibody (Abcam, Cambridge, MA) for 1 h at room temperature. After 3x 5 min washing with PBS, slides were incubated with Cy3-conjugated goat anti-rabbit secondary antibody (Jackson ImmunoResearch Laboratories, Inc., West Grove, PA) for 1 hour at room temperature. After staining, slides were mounted with VECTASHIELD mounting medium (Vector Laboratories, Burlingame, CA).

**CD31 Staining.** The tumor tissue slices were obtained from tumor bearing mice injected with free AF488 dye and AF488 labeled SWNT (SWNT-AF488) solutions at 4h p.i. Frozen tumor tissue slices were taken out from freezer and warmed for 20min at room temperature. Slides were blocked with 10% goat serum in PBS for 15 min at room temperature and then incubated with rat anti-mouse CD31 antibody (BD bioscience, San Jose, CA) for 1 hour at room temperature. After 3x 5 min washing with PBS, slides were incubated with FITC-conjugated goat anti-rat secondary antibody (Jackson ImmunoResearch Laboratories, Inc., West Grove, PA). After staining, slides were mounted with VECTASHIELD mounting medium (Vector Laboratories, Burlingame, CA).

**Fluorescence imaging of tumor slices**

PEGylated SWNTs were labeled by NHS-Alexa Fluor 488 (Invitrogen) at pH 7.5 for 4 h. Excess dye molecules were removed by filtration. SWNT-AF488 and free AF488 with the same fluorescence intensity normalized by a fluorimeter were injected into 4T1 tumor bearing mice, which were sacrificed at 4 h p.i. Tumor slices were stained with Cy3-anti CD31 antibody to visualize the vasculature and imaged by a Zeiss LSM 510 confocal microscope.

**Pharmacokinetics and Biodistribution studies**

Blood circulation was measured by drawing ~10 µl blood from the tail vein of tumor-free healthy Balb/c mice post injection of $^3$H labeled SWNT-PTX, Taxol® or PEG-PTX. The blood samples were dissolved in a lysis buffer (1% SDS, 1% Triton X-100, 40 mM Tris Acetate, 10 mM EDTA, 10 mM DTT) with brief sonication. Concentration of SWNTs in the blood was measured by a Raman method(18, 22) (see below). For $^3$H-PTX measurement, the blood lysate was decolorized by 0.2 ml of 30% hydrogen peroxide (Aldrich) and the radioactivity was counted by Tri-Carb 2800 TR (Perkin-Elmer) scintillation counter following the vendor's instruction. Raman measurement was done as described in the following section. Blood circulation data were plotted as the blood PTX or SWNT levels (unit: %ID/g) against time p.i. Pharmacokinetic analysis was performed by first-order exponential decay fitting of the blood PTX concentration data with the following equation: Blood concentration = $A \times exp\ (-t/\lambda)$, in which *A* was a constant (initial concentration) and *t* was the time p.i. The pharmacokinetic parameters including volume of distribution, areas under curves (AUC) and circulation half lives are calculated and presented in supplementary table S1.

For the biodistribution study, 4T1 tumor bearing mice (tumor size ~200 mm$^3$) were sacrificed at 2h and 24h post injection of $^3$H labeled SWNT-PTX, Taxol® or PEG-PTX. The organs/tissues were collected and split into two halves for $^3$H-PTX and SWNT biodistribution studies. Majority of food residue and feces in the stomach and intestine was cleaned. For the $^3$H-PTX biodistribution, 50-100mg of tissue was weighed and solubilized in 1 mL of scintillation counting compatible soluene-350 solvent (Perkin-Elmer) by incubation at 60$^o$C overnight and decolorized by 0.2 ml of 30% hydrogen peroxide. The $^3$H radioactivity in each organ/tissue was measured by applying the homogenous organ/tissue solutions to a Perkin-Elmer scintillation counter following the vendor's instruction. Biodistribution of $^3$H-PTX was calculated and normalized to the percentage of injected dose per gram tissue (%ID/g). Note that all the biodistribution and circulation tests were carried out at the treatment dose (normalized to 5mg/kg of PTX).

For SWNT biodistribution, the organs/tissues were wet-weighed and homogenized in the lysis buffer (same as used in the blood circulation experiment) with a PowerGen homogenizer (Fisher Scientific). After heating at 70$^o$C for ~2 h, clear homogenous tissue solutions were obtained for Raman measurement (see below).(18, 22) A control experiment was done to confirm that this treatment did not affect the Raman intensity of a SWNT solution (data not shown). The biodistribution of SWNTs in various organs of mice was then calculated and plotted in unit of %ID/g.

**Raman spectroscopy measurement and imaging**

A glass capillary tube filled with tissue lysate solution was placed under the objective (20×) of the Raman microscope. As low as 2 µL of solution sample was needed for each measurement. After focusing at the center of the capillary, we recorded the Raman spectrum of the solution (100 mW power with laser spot size of ~ 25 μm$^2$, 10 second collection time). At least 4 spectra were taken for each sample for averaging. The Raman intensity was obtained by integrating the SWNT G-band peak area from 1570 cm$^{-1}$ to 1620 cm$^{-1}$ and averaged over several spectra. SWNT concentration in blood samples or tissue lysate was determined by comparing the raman intensity with a standard calibration curve obtained for SWNTs in lysis buffer with various known concentrations. The linear dependence between Raman intensity of a SWNT solution and concentration allows for accurate measurement of nanotube concentration in aqueous phase.(18, 22) The concentrations of injected SWNT or SWNT-PTX solutions were also determined by the Raman method prior to injections into mice.

To obtain the Raman mapping image of tumor slices for mice injected with SWNT-PTX, 5μm thick paraffin embedded tumor slices were mounted on SiO$_2$ substrate and mapped under a Renishaw micro-Raman microscope with a line-scan model (100 mW laser power, 40 μm × 2 μm laser spot size, 20 pixels each line, 2 second collection time, 20× objective). The SWNT G-band Raman intensity was plotted vs. x, y positions across the liver slice to obtain a Raman image.

**Necropsy, blood chemistry and histology study**

24 days after initiation of treatment, 3 mice from each treatment group (SWNT-PTX and Taxol®) and 2 age-matched female Balb/c control mice were sacrificed by $CO_2$ asphyxiation. Blood was collected via cardiac puncture at time of sacrifice for analysis of serum chemistries by the Diagnostic Laboratory, Veterinary Service Center, Department of Comparative Medicine, Stanford University School of Medicine. Serum chemistries were run on an Express Plus Chemistry Analyzer (Chiron Diagnostics) and electrolytes were measured on a 644 Na/K/Cl Analyzer (CIBA-Corning). A full necropsy was performed and all internal organs were harvested, fixed in 10% neutral buffered formalin, processed routinely into paraffin, sectioned at 4 microns, stained with hematoxylin & eosin (H&E) and examined by light microscopy.

Examined tissues included: liver, kidneys, spleen, heart, salivary gland, lung, trachea, esophagus, thymus, reproductive tract, urinary bladder, eyes, lymph nodes, brain, thyroid gland, adrenal gland, gastrointestinal tract, pancreas, bone marrow, skeletal muscle, nasal cavities, middle ear, vertebrae, spinal cord and peripheral nerves.

**Statistical analysis**

Quantitative data were expressed as mean ± standard deviation. Means were compared using student's t-test. P values < 0.05 were considered statistically significant.

**Results**

As-grown Hipco SWNTs functionalized by PEGylated phospholipid(14, 18) were used, made by sonication of SWNTs in a water solution of phospholipid-PEG and centrifugation to remove large bundles and impurities. The length distribution of the SWNTs was 20-300nm with a mean of ~100 nm (Supplementary Fig. S1).(14, 18) The PEG functionalized SWNTs exhibited excellent stability without agglomeration in various biological media including serum.(14, 18) We used branched PEG chains for functionalization of SWNTs (see Method) to afford more functional amine groups at the PEG termini for efficient drug conjugation.(22) Paclitaxel was conjugated at the 2'-OH position (24) to the terminal amine group of the branched PEG on SWNTs via a cleavable ester bond (see Method), forming a SWNT-PTX conjugate highly soluble and stable in aqueous solutions (Fig.1a). The un-conjugated paclitaxel was removed thoroughly from the SWNT-PTX solution by filtration. The loading of paclitaxel on SWNTs was characterized to be ~150 per SWNT with ~100nm length by radiolabeling method using tritium ($^3$H) labeled paclitaxel and a UV-VIS-NIR optical absorbance (Fig. 1b, see Method). The SWNT-PTX conjugate was found stable in physiological buffers with little drug release within 48 hours (Supplementary information, Fig. S2). In mouse serum, the release of PTX is faster but SWNT-PTX is still stable for hours (Fig.S2), which is much longer than the blood circulation time of SWNT-PTX as described later. In vitro cell toxicity tests performed with a 4T1 murine breast cancer cell line found that SWNT-PTX exhibited similar toxicity as Taxol® and PEGylated PTX (Fig. 1c) without any loss of cancer cell destruction ability. Consistent to the previous studies, (7-14) no noticeable toxic effect to cells was observed for plain nanotube carriers without drug even at high SWNT concentrations (Supplementary information, Fig. S3).

We next moved to the *in vivo* cancer treatment on the paclitaxel resistant 4T1 murine breast cancer mice model.(25, 26) Female Balb/c mice bearing subcutaneously inoculated 4T1 tumors were treated with different forms of paclitaxel over several weeks including the clinical Taxol® formulation, PEGylated PTX (PEG-PTX, see Method), DSEP-PEG conjugated PTX (DSPE-PEG-PTX) and SWNT-PTX (14 mice in this group). The treatments were done by injecting Taxol®, PEG-PTX, DSEP-PEG-PTX and SWNT-PTX (at the same PTX dose of 5mg/kg for all three formulations, once every 6 days) intravenously into tumor-bearing mice. The mice were observed daily for clinical symptoms and the tumor volume was measured by a digital caliper every other day. As shown in Fig. 2, a time-related increase in tumor volume was observed in the control untreated group and SWNT vehicle only group in which the tumors showed average fractional tumor volumes ($V/V_0$) of 10.1 ± 1.7 and 9.8 ± 2.0, respectively on day 22. Taxol® treatment, PEG-PTX treatment and DSPE-PEG-PTX treatment resulted in $V/V_0$ of 7.3 ± 1.5 (P = 0.06 vs untreated), 8.0 ± 1.6 (P = 0.18 vs untreated), 8.6 ± 0.9 (P = 0.33 vs untreated) on day 22, which represents tumor growth inhibition (TGI) of 27.7%, 20.8% and 14.9% respectively. In contrast, SWNT-PTX treatment resulted in a $V/V_0$ of 4.1 ± 1.1 on day 22 (P = $2.4 \times 10^{-6}$ vs untreated, P = 0.00063 vs Taxol®, P = 0.00026 vs PEG-PTX, $2.7 \times 10^{-5}$ vs DSEP-PEG-PTX), representing a TGI of 59.4 %, which is significantly more effective than Taxol®, PEG-PTX and DSPE-PEG-PTX.

To investigate the tumor suppression mechanism, we performed terminal transferase dUTP nick end labeling (TUNEL) assay to examine the apoptosis level in the tumors(27) from mice received different treatments. Similar to untreated tumor, Taxol® treated tumor showed only 2-3% of apoptotic cells (Fig. 3a 1st and 2nd rows, supplementary Fig. S4a). In contrast, high apoptosis level (~70%, P<0.0001 vs untreated and Taxol® treated tumors) was observed in SWNT-PTX treated tumor (Fig. 3a, 4th row, and see supplementary Fig. S4a for quantitative comparison), consistent with the improved tumor growth inhibition efficacy (Fig.2). The Ki-67 antibody staining method has been widely used as a cell proliferation marker to stain proliferation active cells in the G1, G2 and S phases of cell cycle.(28) We found that cell proliferation in Taxol® treated tumor was as active as in untreated tumor (Fig. 3b 2nd row, see supplementary Fig. S4b for quantitative comparison). In the SWNT-PTX treated tumor case however, only ~20% of proliferation active cells were noted compared with the number in the untreated tumor (Fig. 3b 3rd row, Fig. S4b, P<0.0001 vs untreated and Taxol® treated tumors). As the control, plain SWNT without PTX showed no effect to the tumors (Fig. 3, 3rd row), proving the treatment efficacy of SWNT-PTX is due to PTX carried into tumors by nanotubes. Thus, both TUNEL staining and Ki67 staining results clearly confirmed the treatment efficacy of SWNT-PTX by inhibiting proliferation and inducing apoptosis of tumor cells.

To investigate the pharmacokinetics of various drug complexes, we first measured blood circulation behaviors of PEGylated SWNTs with and without PTX conjugation by Raman spectroscopic detection of SWNTs in blood sample drawn from mice post injection (p.i.) of SWNT and SWNT-PTX (see Method). We observed a significantly shortened circulation half life of our branch-PEGylated SWNT from ~3.3 h to ~1.1 h (circulation half life was obtained by one compartment first order exponential decay fitting, see Method) after PTX conjugation (Fig. 4a). This result was important and attributed to the high hydrophobicity of conjugated paclitaxel, reducing the biological inertness of the PEGylated nanotubes in vivo and shortening the blood circulation time. Blood circulation behaviors of the three forms of PTX were measured using $^3$H labeled paclitaxel. Liquid scintillation counting of $^3$H-PTX radioactivity of blood samples collected from mice post-injection showed circulation half lives of 18.8 ± 1.5 min, 22.8 ± 1.0 min and 81.4 ± 7.4 min for $^3$H-PTX injected in Taxol®, PEG-PTX and SWNT-PTX respectively (Fig. 4b, see supplementary table S1 for complete pharmacokinetic data). This clearly revealed that conjugation of PTX to PEGylated SWNTs significantly increased the blood circulation time of PTX. Interestingly, simple PEGylation of PTX, though imparted water solubility of PTX, still exhibited much shorted blood circulation than PTX on PEGylated SWNTs. Note that for SWNT-PTX, circulation curves of radiolabeled PTX measured by radioactivity (Fig.4a&b green curves) and the drug carrier SWNT measured by Raman have consistent slopes (Fig. 4a red curve), suggesting that PTX and SWNT remained in a conjugated form in the blood circulation stage, which is consistent to the relatively slow PTX releasing behavior of SWNT-PTX in mouse serum (Supplementary information, Fig. S2). The minor difference in the absolute values could be due to systematic errors between two different methodologies.

To understand the tumor treatment efficacy of various PTX formulations, i.e., SWNT-PTX, Taxol® and PEG-PTX, we investigated biodistribution of $^3$H-PTX in the tumor and various main organs. We observed significant differences in the biodistribution of PTX administrated in the three formulations of PTX (Fig.4c&d). Consistent with the blood circulation data (Fig. 4b), SWNT-PTX showed noticeable PTX activity in blood at 2h p.i., while PTX levels in the blood were much lower in the Taxol® ($P < 0.001$) and PEG-PTX ($P < 0.01$) cases (Fig. 4c insert). Differences in biodistributions of PTX in the three cases were the most obvious at 2h p.i., with much higher PTX signals in the RES organs (liver/spleen) and intestine of mice in the SWNT-PTX case than the two other cases (Fig.4c).

Importantly, SWNT-PTX afforded much higher PTX uptake in the tumor than Taxol® and PEG-PTX. The tumor PTX levels in the SWNT-PTX case was higher than those of Taxol® and PEG-PTX by 10 and 6-fold respectively at 2h p.i. (Fig. 4c), and by 6 and 4-fold higher respectively at 24 h p.i. (Fig. 4d, P<0.001 in all cases). The ability of higher drug delivery efficiency to tumor by our PEGylated SWNTs was striking and directly responsible for the higher tumor suppression efficacy of SWNT-PTX than the other formulations. This suggests that to reach similar tumor uptake of drug, much lower injected dose can be used by SWNT delivery than Taxol®, which is highly favorable for lowering toxic side effect to normal organs and tissues. An important gauge to drug delivery efficiency is the tumor-to-normal organ/tissue PTX uptake ratios (T/N ratios). We obtained significantly higher T/N PTX uptake ratios (for tumor over liver, spleen, muscle and other organs examined) in the case of SWNT-PTX than Taxol® and PEG-PTX (except at 2 h p.i. for spleen) at the 2 h and 24 h (Fig. 4e&f, supplementary table S2). This again makes SWNT-PTX highly favorable for high tumor suppression efficacy and low side effects.

We investigated the biodistribution of SWNTs injected as SWNT-PTX conjugates into mice by utilizing their intrinsic Raman scattering properties without relying on radio or fluorescent labels.(18, 29) We observed high uptake of SWNTs in the reticuloendothelial systems (RES)(18-20) including liver and spleen (Fig. 5 a-c). Tumor uptake of SWNT-PTX increased significantly from ~1%ID/g at 30 min to ~5%ID/g at 2h, indicating accumulation of SWNT-PTX during this period through blood circulation (see Fig.4b for circulation curve). Tumor uptake of SWNTs at 4.7% (std. = 2.1%, n = 3) ID/g was observed at 2h p.i. (Fig.5b), reasonably consistent with the ~6.4% (std. = 1.1%, n = 3) ID/g PTX tumor uptake (Fig.5b), suggesting that SWNT-PTX was taken up by tumor in a conjugated form. The SWNT biodistribution exhibited little change from 2h (Fig.5b) to 24h p.i. (Fig.5c), in contrast to the biodistribution of radiolabeled PTX (Fig.4c vs. Fig.4d green bars). This suggests that the dissociation of PTX from SWNT carriers in vivo resulted from in vivo cleavage of the ester bond between SWNT and PTX is likely by carboxylesterases.(30-32)

We carried out micro-raman imaging of SWNTs in tumor slices upon sacrificing mice treated by SWNT-PTX at 24 h p.i. The tumor uptake of SWNTs was indeed confirmed by Raman mapping of the SWNT characteristic G-band Raman peak at ~1580 cm$^{-1}$ in the tumor with a spatial resolution of ~ 1μm (Fig. 5c, inset). To investigate the location of nanotubes in the tumor relative to the vasculature, we injected Alexa Fluor 488 (AF488) fluorescently labeled SWNTs into 4T1 tumor bearing mice, sacrificed the mice, collected the tumors for vasculature staining and fluorescence imaging (Fig. 5e). We observed fluorescently labeled SWNTs both with and without overlaying with tumor vasculatures (Fig. 5e, lower right image). This suggested that while most SWNTs appeared to be located in or near the tumor vasculature, a fraction of nanotubes could leak through the tumor vessel into the tumor interstitial space. As the control, no tumor retention of fluorescent dye was observed in mice injected with free AF488 at the same dose (Fig. 5d).

Toxic side effects to normal organs and overall well being have been the main problems of cancer chemotherapeutics. By themselves, our well PEGylated SWNTs have been found to be non-toxic to mice in vivo monitored over many months.(21, 22) We carried out a pilot toxicity study by treating healthy, tumor-free Balb/c mice with Taxol® and SWNT-PTX at the same 5mg/kg PTX dose once every six days. We observed neither mortality nor noticeable body weight loss of the mice treated with SWNT-PTX and Taxol® compared to untreated control group at this relatively low PTX dose and injection frequency (Fig. 6a). Blood chemistry test was performed 24 days after initiation of the treatment, showing no physiologically significant difference among the 3 groups (Fig. 6b & Supplementary Table S3). Furthermore, hematoxylin & eosin (H&E) stained sections of the 25 organs and organ systems were examined (Fig.6c), without noticing obvious abnormal damage in the main organs including the liver and spleen that had high SWNT uptake, which was consistent to the normal hepatic enzyme levels measured in the blood chemistry test (Fig. 6b). The observed lack of obvious toxic side effect was partly due to the low dose of PTX used as the maximum tolerable dose of PTX in the Taxol® case ~20-50mg/kg.(33-35) Achieving tumor treatment efficacy by SWNT-PTX at a PTX dose well below the toxic limit is owed to ability of drug delivery to tumors by SWNTs. However, further careful studies such as the hepatic macrophage function tests are required to examine any potential near-term or long-term side effect our SWNT-PTX.

## Discussion

We have shown that SWNT delivery of PTX affords markedly improved treatment efficacy over clinical Taxol®, evidenced by its ability of slowing down tumor growth at a low PTX dose. The treatment effect is confirmed by tumor staining that reveals significant apoptotic cells and few proliferation active cells in the SWNT-PTX treated tumor. The key reason for higher tumor suppression efficacy of SWNT-PTX than Taxol® and PEG-PTX is the up to 10-fold higher tumor uptake of PTX afforded by SWNT carriers, which is a remarkable result. This is directly responsible for tumor suppression at a low dose of SWNT-PTX for the 4T1 tumor model normally resistant to PTX treatment.(25)

Prolonged blood circulation and EPR effects are responsible for significantly higher tumor uptake of PTX in the SWNT-PTX case (6.4 %ID/g at 2h p.i.) than Taxol® (0.6 %ID/g) and PEG-PTX (1.1 %ID/g). The poor water solubility of various cancer therapeutic drugs limits their clinical applications. Cremophor EL is a commonly used reagent to disperse paclitaxel and other drugs in saline for administration. However, its toxic effects have been noted in both animal models and patients.(36-39) Similar to previous reports,(40-45) we observe (Fig. 4b) short blood circulation time for PTX in Taxol®. Little PTX (<2 %ID/g) in the Taxol® form remains circulating in the blood after only 11 min post injection at the current 5mg/kg injected dose (Fig.4b). PTX in Taxol® is known to be cleared from the blood and taken up by various organs especially kidney and liver for rapid renal and fecal excretion with very low tumor uptake. (40-42, 44)

Branched PEGylation of PTX via similar ester linkage as in SWNT-PTX conjugates affords water solubility of PTX. However, the blood circulation time is still short (PEG-PTX concentration diminished to <2%ID/g in ~70 min p.i.), albeit longer than Taxol®. PEG-PTX remains a relatively small molecule that tends to be rapidly excreted via the kidney and renal route, evidenced by the high kidney and urine signals of radiolabeled PEG-PTX (data not shown). This leads to little advantage of PEGylation of PTX over Taxol® in tumor uptake and treatment efficacy, as found here and by previous PTX PEGylation work.(46)

The water solubility of our SWNT-PTX formulation favors prolonged blood circulation. Nevertheless, the high hydrophobicity of PTX reduces the hydrophilicity and biological inertness of our branch-PEG functionalized SWNTs, causing significantly shortened blood circulation half lives of the SWNT-PTX formulation (~1.1 h) compared to PEGylated SWNTs without PTX attachment (~3.3 h) (Fig. 4a). Compared to PEG-PTX, SWNT-PTX exhibit finite lengths (20-300nm, mean~100nm, Supplementary Fig. S1), a factor that favors long blood circulation since the average length of the nanotubes exceeds the threshold for renal

clearance.(47) Pharmacokinetics of materials with long blood circulation times are typically desired for a drug delivery vehicle for tumor treatment (2, 48, 49) to favor high tumor accumulation from the circulating blood through EPR effects. Note that our method of drug delivery by PEGylated SWNTs should be readily applicable to a wide range of hydrophobic or water-insoluble drugs. This could lead to a general drug delivery strategy for potent but water insoluble molecules.

Tumor staining data clearly revealed apoptotic cells (Fig. 3a) inside the tumor treated by SWNT-PTX. Nanotubes were observed in the tumor vasculature as well as leaked out of the vessels (Fig. 5e). Drug delivery to cancer cells through the tumor vessel walls and interstitial space is desired for high tumor treatment efficacy. SWNTs appeared to exhibit certain ability in overcoming these barriers, which could be related to the quasi 1D shape of these materials. An interesting feature of SWNTs is that the length of nanotubes (20-300 nm currently) could be controlled more precisely to span various size regimes. This could allow for investigation of length effect of 1D materials to the tumor penetration and suppression efficacy of drug complexes. This intriguing length effect will require systematic exploration in the future.

PTX conjugation to PEGylated SWNTs clearly alters the pharmacokinetics and biodistribution of PTX from Taxol® and PEG-PTX. The up to 10X high tumor uptake of the drug through SWNT-PTX and high T/N ratios, strongly favor high tumor killing efficacy and low toxicity to normal organs. High RES uptake is known for nanomaterials in general. The high uptake of SWNT-PTX in RES organs such as liver and spleen(18, 22) could be a cause of concern in terms of toxicity to these organs. Importantly, our biodistribution studies revealed relatively low PTX levels in the RES organs at later time points (2 h and 24 h), differing from the SWNT biodistribution (Fig.5a-c). The difference between the biodistribution of SWNT and $^3$H-PTX (measured by radioactivity) suggests rapid release of PTX from SWNT in the various organs and tissues in vivo, resulted from in vivo cleavage of the ester linkage between PTX and PEGylated SWNT most likely by carboxylesterases especially those in the liver.(30-32)  We observed a significant PTX accumulation but proportionally lower SWNT signal in the intestine in the PTX-SWNT case at 30 min and 2 h p.i.(Fig. 5a&b). We also detected strong PTX signal in the feces even at only 30 min p.i.(data not shown). These data suggested that SWNT-PTX taken up by the RES organs were dissociated via ester cleavage for release for excretion. Unlike the SWNT carriers, which are excreted gradually excreted in weeks or even months,(22) the dissociated PTX drug molecules can be rapidly excreted via both feces and urine without causing noticeable toxicity. Taken together, the uptake of drug-nanomaterial complexes by RES could serve as a scavenger system to eliminate toxic drugs as well as carriers.

The maximum tolerable dose of Taxol® for Balb/c mice is reported to be in the range of 20-50mg/kg.(33-35) Achieving tumor growth suppression by SWNT-PTX at 5mg/kg dose once every six days suggests the promise of SWNT drug delivery for effective cancer treatment with low side effects. More importantly, our water-soluble SWNT-PTX formulation is cremophor free. SWNTs have been proved to be safe at least in mouse models. (21, 22) The amount of SWNTs required to give 5mg/kg of PTX is only ~4mg/kg, compared with ~420mg/kg cremophor in the Taxol® case for the same PTX dose. Further, the same SWNT conjugation strategy applies to many other water-insoluble drugs. Although its treatment efficacy and side effects need to be carefully compared with the latest FDA approved palcitaxel formulation, Abraxane, our preliminary results indicate comparable or even better efficacy of SWNT-PTX than Abraxane (data not shown).

SWNTs are highly promising for drug delivery due to several factors. These materials can now be functionalized to a sufficient degree to facilitate nearly complete excretion of SWNTs from mice over time.(22) The chemical composition (purely carbon) of carbon nanotubes is among the safest in the inorganic nanomaterials, many of which such as quantum dots have heavy metal compositions. The unique 1D structure and tunable length provide an ideal platform to investigate size and shape effects in vivo. Lastly, unlike the conventional organic drug carriers, the intrinsic spectroscopic properties of nanotubes including Raman and photoluminescence can provide valuable means of tracking, detecting and imaging to understand the in vivo behavior and drug delivery efficacy in vivo. Taken together, carbon nanotubes are promising materials for potential multimodality cancer therapy and imaging.

To our knowledge, this is the first successful report that carbon nanotubes are used as drug delivery vehicles to achieve in vivo tumor treatment efficacy with mice. This opens up further exploration of biomedical applications of novel carbon nanomaterials with animals for potential translation into the clinic in the future. The treatment efficacy of SWNT based drug delivery vehicles could be further improved by optimization of the surface chemistry and size of nanotubes as well as the positioning of drug molecules for desired pharmacokinetics. Targeting ligands on nanotubes for tumor targeted drug delivery is also expected to further enhance treatment efficacy. The 1D shape and length of nanotubes easily allow for targeting ligands, drugs and multiple molecules for synergistic effects.

**Acknowledgements**

This work was supported by NIH-NCI CCNE-TR at Stanford (H.D.), a Stanford Bio-X Initiative Grant and a Stanford Graduate Fellowship.

All authors discussed the results and commented on the manuscript.

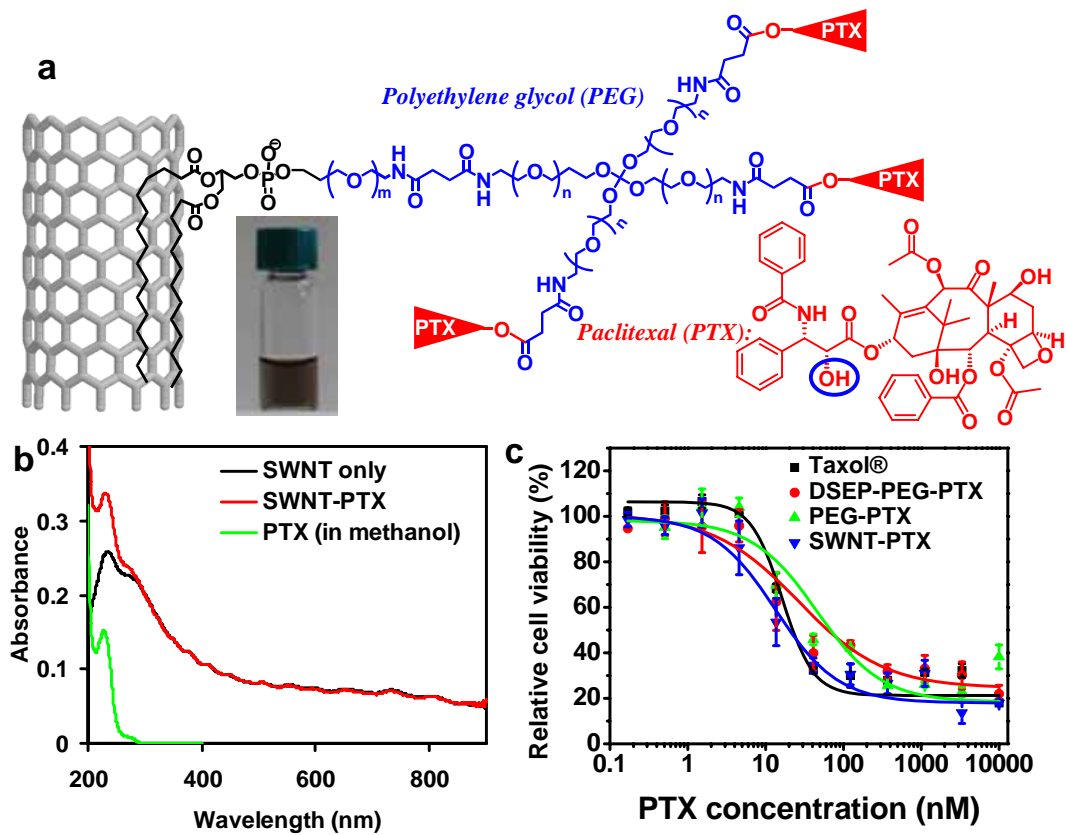

**Figure 1.** Carbon nanotube for paclitaxel delivery. **a,** schematic illustration of paclitaxel conjugation to SWNT functionalized by phospholipids with branched-PEG chains. The PTX molecules are reacted with succinic anhydride (at the circled OH site) to form cleavable ester bonds and linked to the termini of branched PEG, via amide bonds. This allows for releasing of PTX from nanotubes by ester cleavage in vivo. The SWNT-PTX conjugate is stably suspended in normal physiological buffer (PBS, as shown in the photo) and serum without aggregation. **b,** UV-VIS-NIR spectra of SWNT before (black curve) and after PTX conjugation (red). The absorbance peak of PTX at 230nm (green curve) was used to measure the PTX loading on nanotubes and the result was confirmed by radiolabel based assay. Excess un-conjugated PTX was removed by extensive filtration and washing. **c,** cell survival vs. concentration of PTX for 4T1 cells treated with Taxol®, PEG-PTX, DSEP-PEG-PTX or SWNT-PTX for 3 days. The PTX concentrations to cause 50% cell viability inhibition (IC50 values) were determined by sigmoidal fitting to be 16.4 ± 1.7 nM for Taxol®, 23.5 ± 1.1 nM for DSPE-PEG-PTX, 28.4 ± 3.4 nM for PEG-PTX and 13.4 ± 1.8 nM for SWNT-PTX. Error bars based on four parallel samples. Plain SWNTs (no PTX conjugated) are non-toxic (see supplementary Fig.S3)

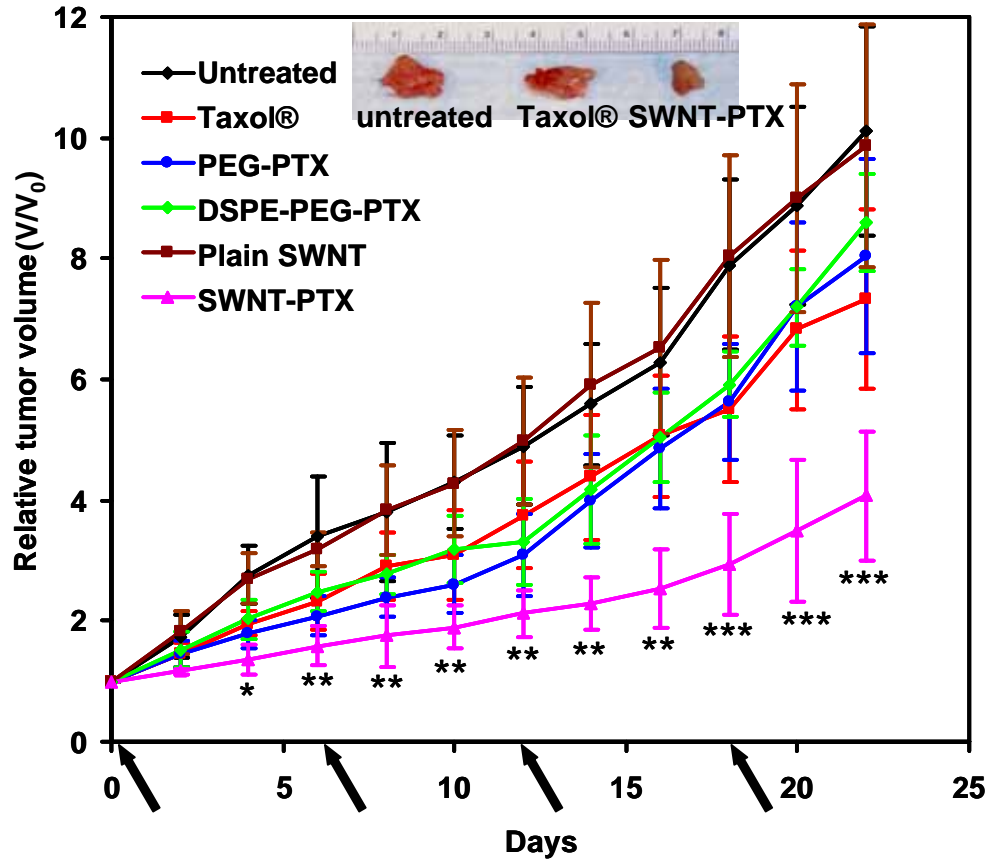

**Figure 2.** Nanotube paclitaxel delivery suppresses tumor growth 4T1 breast cancer mice model. **a,** Tumor growth curves of 4T1 tumor bearing mice received different treatments indicated. The same PTX dose (5 mg/kg) was injected (on day 0, 6, 12 and 18, marked by arrows) for Taxol®, PEG-PTX, DSEP-PEG-PTX and SWNT-PTX. P values (Taxol® vs SWNT-PTX): * $p<0.05$, ** $p<0.01$, *** $p<0.001$. Number of mice used in experiments: 8 mice per group for untreated, 5 mice per group for SWNT only, 9 mice per group for Taxol®, 5 mice per group for PEG-PTX, 6 mice per group for DSEP-PEG-PTX, 14 mice per group for SWNT-PTX. Inset: a photo of representative tumors taken out of an untreated mouse (left), a Taxol® treated mouse (middle) and a SWNT-PTX treated mouse after sacrificing the mice at the end of the treatments.

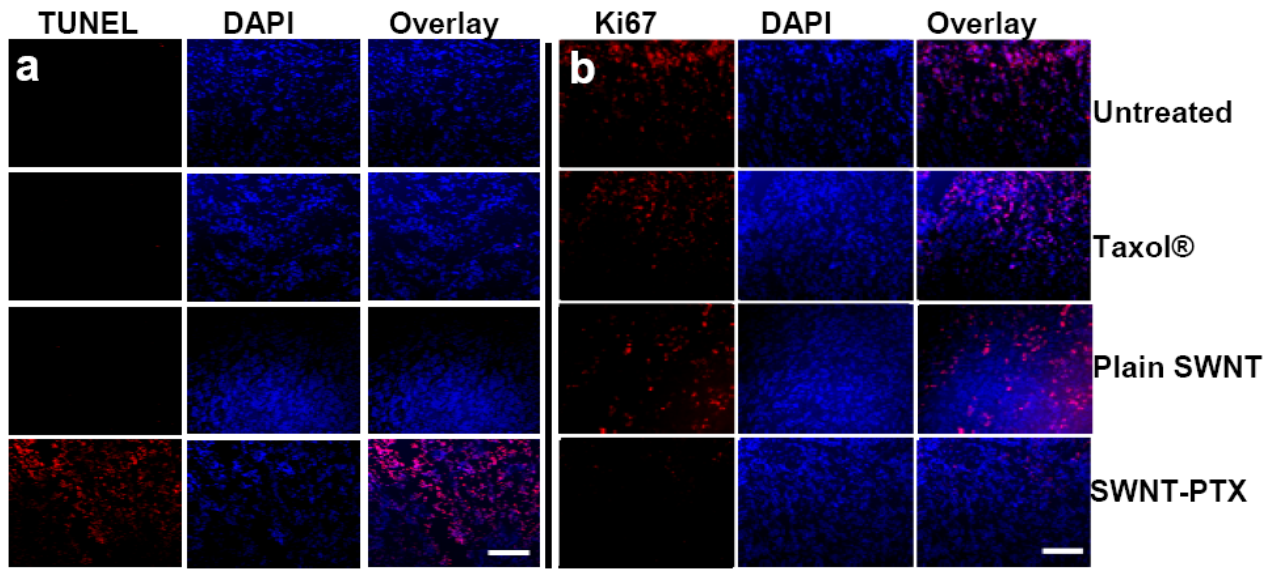

**Figure 3**. Tumor staining for understanding of treatment effects. **a,** TUNEL (apoptosis assay) and DAPI (nuclear) co-staining images of 4T1 tumor slices from mice after different treatments indicated. While tumors from untreated mice (1st row), Taxol® treated mice (2nd row) and plain SWNT treated mice (3rd showed few apoptotic cells, many cells in the tumor from SWNT-PTX treated mice (4th row) were undergoing apoptosis. **b,** Ki67 (proliferation assay) and DAPI co-staining images of tumor slices from mice after various treatments. Few proliferation active cells were observed in the tumor received SWNT-PTX treatment (4th row). Tumors used in this study were taken from 4T1 tumor bearing mice 12 days post initiation of treatment. Scale bar: 100 µm.

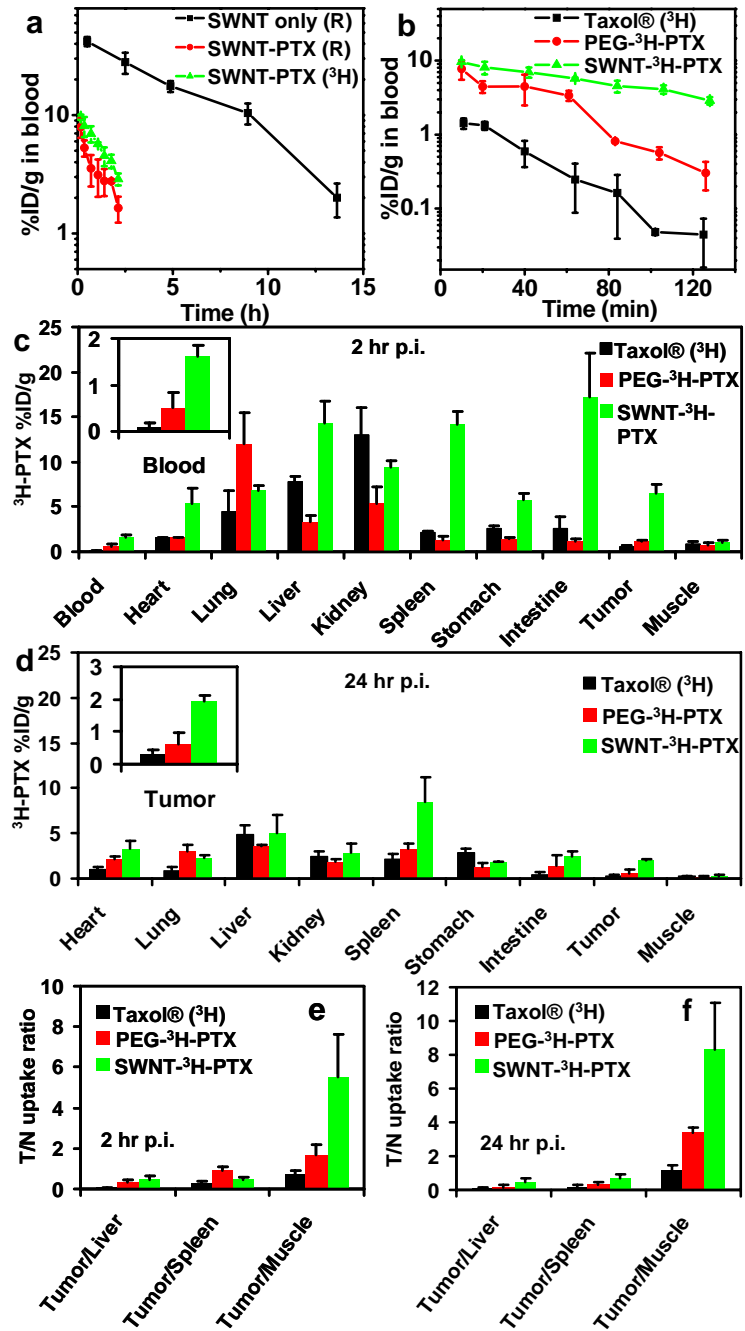

**Figure 4.** Pharmacokinetics and biodistribution. **a,** blood circulation data of SWNT with and without PTX conjugation (marked as *SWNT-PTX (R)* and *SWNT only (R)* respectively) measured by Raman detection of SWNTs in blood samples (see Methods). Blood circulation data for SWNT-$^3$HPTX (green curve) was also obtained by scintillation counting of $^3$H radioactivity in blood. Conjugation of PTX onto SWNTs greatly shortened circulation half life of SWNTs from 3.3 to 1.1 h. **b,** blood circulation data of $^3$H labeled Taxol®, PEG-PTX and SWNT-PTX measured by scintillation counting. SWNT-PTX exhibited significantly prolonged circulation half life (81.4 ± 7.4 min) than that of Taxol® (18.8 ± 1.5 min) and PEG-PTX (22.8 ± 1.0 min). **c & d,** $^3$H-PTX biodistribution in 4T1 tumor bearing mice injected with $^3$H labeled Taxol®, PEG-PTX and SWNT-PTX at **(c)** 2 h p.i. and **(d)** 24 h p.i. Inset in **(c)**: $^3$H-PTX levels in the blood at 2 h p.i. Inset in **(d)**: $^3$H-PTX levels in the tumor at 24 h p.i. **e & f,** tumor to normal organ/tissue ratios (T/N ratios) of the 3 formulations of PTX at **(e)** 2 h and **(f)** 24 h p.i. SWNT-PTX exhibited the highest T/N ratio among the 3 PTX formulations, even when compared with the RES organs (except for spleen at 2 h p.i.) which had dominant SWNT uptake (see Fig. 5). The error bars were based on 3 mice per group in all graphs. 5mg/kg PTX dose was used in all cases.

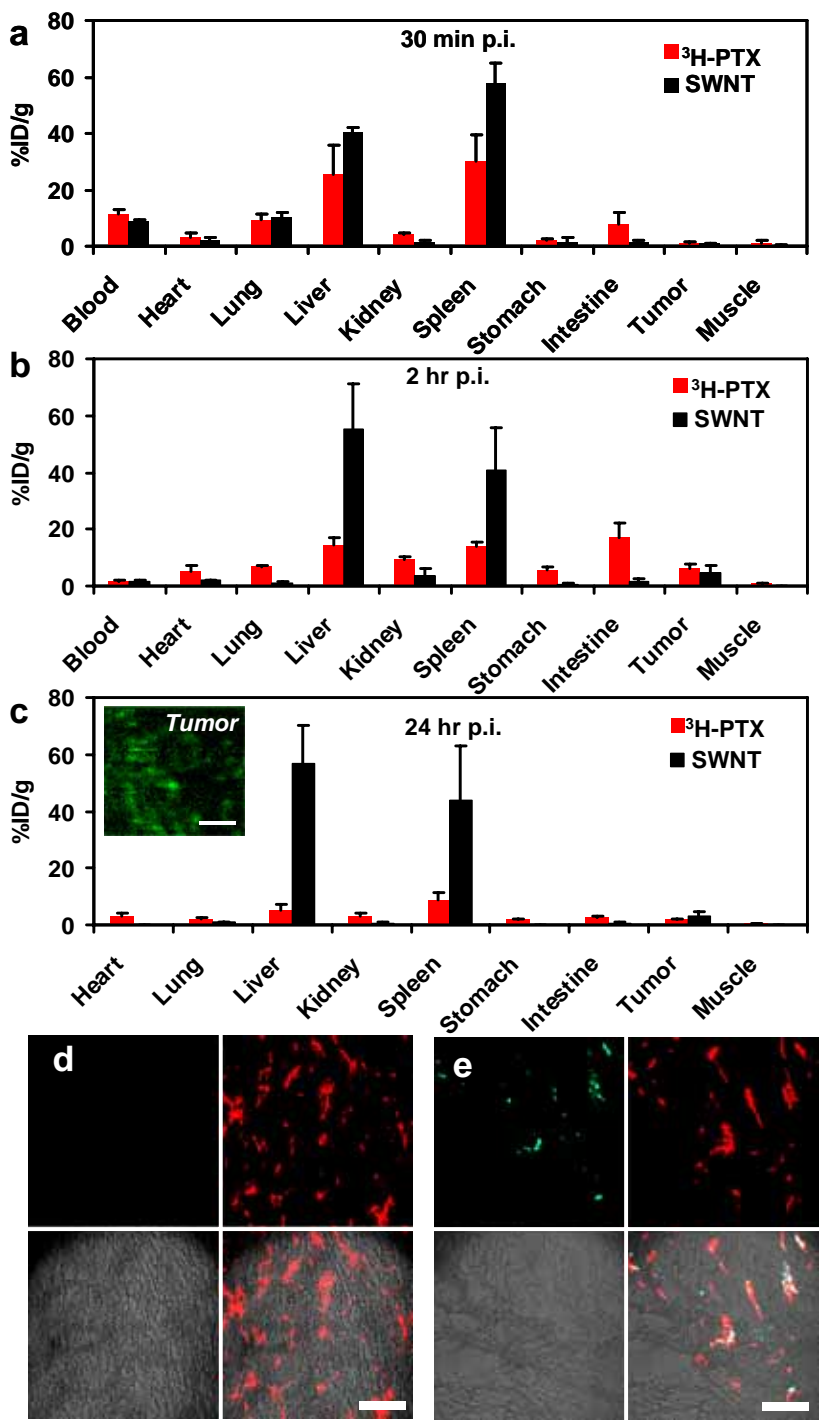

**Figure 5.** SWNT biodistribution measured by Raman spectroscopy. **a - c,** comparison of $^3$H-PTX biodistribution and SWNT biodistribution in mice injected with SWNT-PTX($^3$H) at **(a)** 30 min, **(b)** 2 h and **(c)** 24 h p.i. SWNT biodistribution was measured by a Raman method (see Methods). The different biodistributions of PTX and SWNT carrier suggest rapid cleavage of ester bond for releasing of PTX from SWNTs in vivo. Error bars in all graphs were based on 3 mice per group. Insert in **(c)**: a Raman image of the tumor slice. Strong SWNT G-band Raman signals at ~1580cm$^{-1}$ shift (green color corresponds to high G-band intensity) were observed in the tumor. Scale bar: 50 μm. **d & e,** confocal fluorescence images of tumor slices from mice injected with **(d)** free AF488 dye and **(e)** AF488 labeled SWNT (SWNT-AF488). Tumor vasculature was stained by Cy3-anti-CD31 (red color). AF488 fluorescence (green color) and vasculature fluorescence were overlaid with optical image (bottom left) to obtain the bottom right images. Scale bar: 100μm.

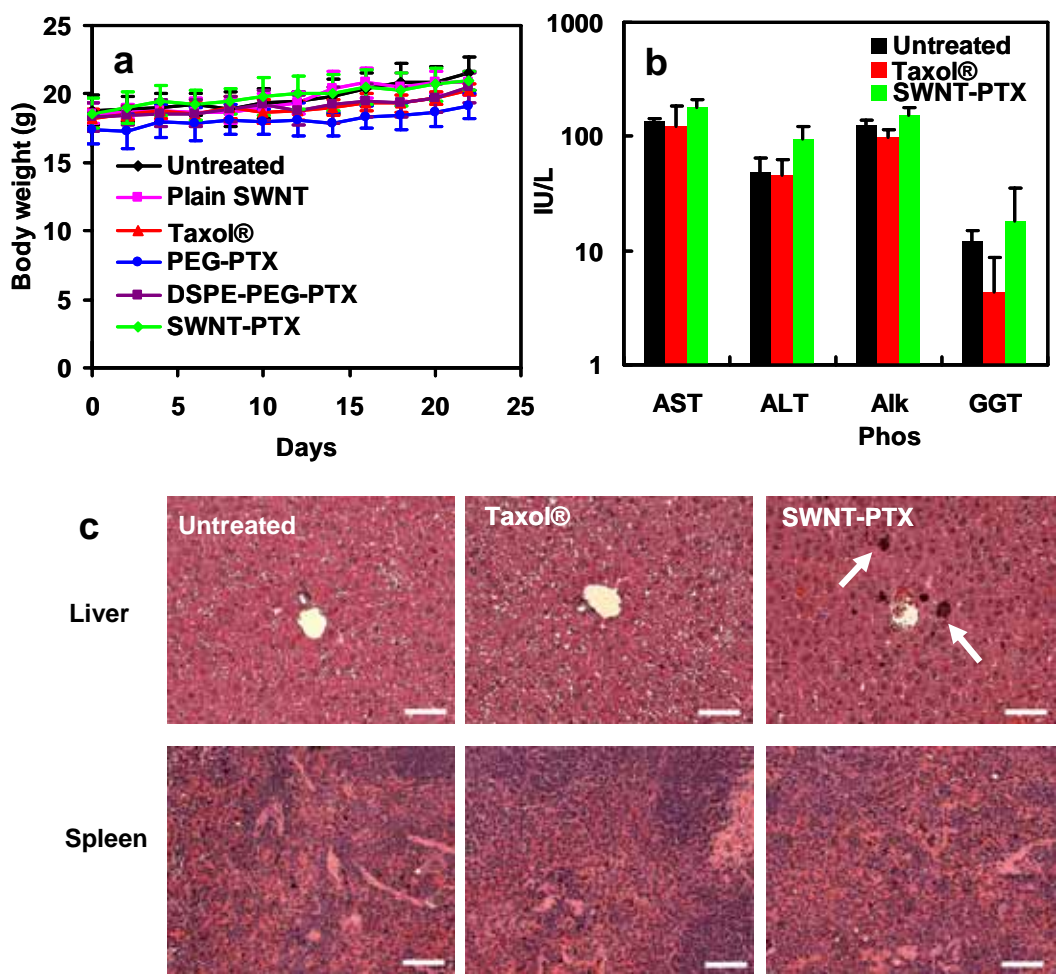

**Figure 6.** Pilot toxicity study. **a,** body weight curves of mice received different treatments in the study (PTX dose~5mg/kg). No obvious loss of body weight was observed in all the groups. 5-14 mice were used in each group (see details in Fig. 2 caption). **b,** blood chemistry data of untreated, Taxol® treated and SWNT-PTX treated mice. Specific attention was paid to those hepatic related serum chemistries (which would reflect liver damage or alternation of function) including aspartate aminotransferase (AST), alanine transaminase (ACT), alkaline phosphatase (Alk Phos) and gamma-glutamyl transpeptidase (GGT), without finding obvious abnormality for SWNT-PTX treated mice. The error bars are based on 3 mice in each group. **c,** hematoxylin & eosin (H&E) stained liver and spleen slices of mice. Although residues of carbon nanotubes were observed as black dots in the liver as pointed by the white arrow, no obvious damage was noticed in the liver and spleen of SWNT-PTX treated mice. Scale bar: 50 µm.